# Fermi level depinning *via* insertion of a graphene buffer layer at the gold–2D tin monoxide contact

**Yujia Tian[1,2], Devesh R. Kripalani[1], Ming Xue[2], and Kun Zhou[1*]**

[1]School of Mechanical and Aerospace Engineering, Nanyang Technological University, Singapore 639798, Singapore

[2]Infineon Technologies Asia Pacific Pte. Ltd., Singapore 349282, Singapore

*Email: kzhou@ntu.edu.sg

**ORCID:**

Y. Tian: 0000-0002-6032-4715

D. R. Kripalani: 0000-0002-1430-1133

K. Zhou: 0000-0001-7660-2911

# ABSTRACT

Two-dimensional (2D) tin monoxide (SnO) has attracted much attention owing to its distinctive electronic and optical properties, which render itself suitable as a channel material in field effect transistors (FETs). However, upon contact with metals for such applications, the Fermi level pinning effect may occur, where states are induced in its band gap by the metal, hindering its intrinsic semiconducting properties. We propose the insertion of graphene at the contact interface to alleviate the metal-induced gap states. By using gold (Au) as the electrode material and monolayer SnO (mSnO) as the channel material, the geometry, bonding strength, charge transfer and tunnel barriers of charges, and electronic properties including the work function, band structure, density of states, and Schottky barriers are thoroughly investigated using first-principles calculations for the structures with and without graphene to reveal the contact behaviours and Fermi level depinning mechanism. It has been demonstrated that strong covalent bonding is formed between gold and mSnO, while the graphene interlayer forms weak van der Waals interaction with both materials, which minimises the perturbance to the band structure of mSnO. The effects of out-of-plane compression are also analysed to assess the performance of the contact under mechanical deformation, and a feasible fabrication route for the heterostructure with graphene is proposed. This work systematically explores the properties of the Au–mSnO contact for applications in FETs and provides thorough guidance for future exploitation of 2D materials in various electronic applications and for selection of buffer layers to improve metal–semiconductor contact.

# 1. INTRODUCTION

Field effect transistors (FETs) have been successfully miniaturised over the past few years, but scaling silicon-based FETs to channel lengths below 10 nm is extremely difficult because of serious short-channel effects and source-to-drain tunnelling [1, 2]. To solve the problem, subnanometer monolayer FETs have been successfully fabricated, with remarkable on/off ratios [3-5]. In particular, two-dimensional (2D) tin monoxide (SnO) has garnered much research interest owing to its distinctive electronic and optical properties for a wide range of technological applications [6]. It is a semiconductor exhibiting native p-type conductivity [7]. SnO is stable, environmentally friendly, abundant, and naturally occurring as the mineral Romarchite [8]. These unique properties have enabled 2D SnO to be suitable for use as a channel material in FETs [9, 10], and it has been successfully fabricated [8, 11].

Actual devices based on monolayer SnO (mSnO) will have the metal–mSnO contact. In the 2D limit, the properties of the interface depend on the interaction between the metal and the semiconductor, which governs their performance and applications. Strong interface interaction may induce the Fermi level pinning (FLP) effect with localised density of states (DOS) at the interface [12]. These metal-induced gap states (MIGS) originate from a decaying metallic wave function into the nanometer depth of semiconductors [13] and serve as reservoirs for charges, hence pinning the Fermi level $E_f$. The FLP effect hinders the tailoring of the Schottky barrier height (SBH) [14-16]. A low SBH at the source or drain contact is essential to achieving high drive current in the FET.

Breaking the interface interaction by introducing vacuum spacing at the interface is not practical during the manufacturing of the metal–semiconductor contact. But inserting an inert layer between the two materials can be [17-21], and this layer needs to be thin enough to facilitate charge transfer across the interface. Being an extremely inert 2D material with no band gap, graphene seems to be an ideal candidate, and atomically clean metal–graphene–n-type silicon junctions have been successfully fabricated to increase the current density at reverse bias by minimising the MIGS [20].

Going beyond the analytical Schottky barrier (SB) theory, we present a systematic first-principles study of the contact between gold (Au), a common electrode in experiments [3] and mSnO. The geometry, bonding strength, charge transfer and tunnel barriers of charges, as well as electronic properties including the work function, band structure, DOS, and Schottky barriers (both vertical and lateral) are thoroughly investigated and compared for Au–mSnO and Au–graphene–mSnO structures to reveal their contact properties and the mechanism of Fermi level depinning by graphene insertion. To facilitate the application of mSnO in monolayer FETs, the effects of out-of-plane compression are also analysed to simulate the unavoidable mechanical deformation during their operation, test, and transport. We also propose a fabrication order for the heterostructure with the interlayer based on the stability analysis.

## 2. COMPUTATIONAL DETAILS

**Supercell construction**

In 2D material–bulk metal systems, 4–6 layers of metal atoms are usually used to model the bulk metal [22-24]. A slab model consisting of five atomic layers is chosen here to imitate the metal gold in our Au–mSnO supercell. This number is decreased to four in our Au–graphene–mSnO supercell to reduce the computation time.

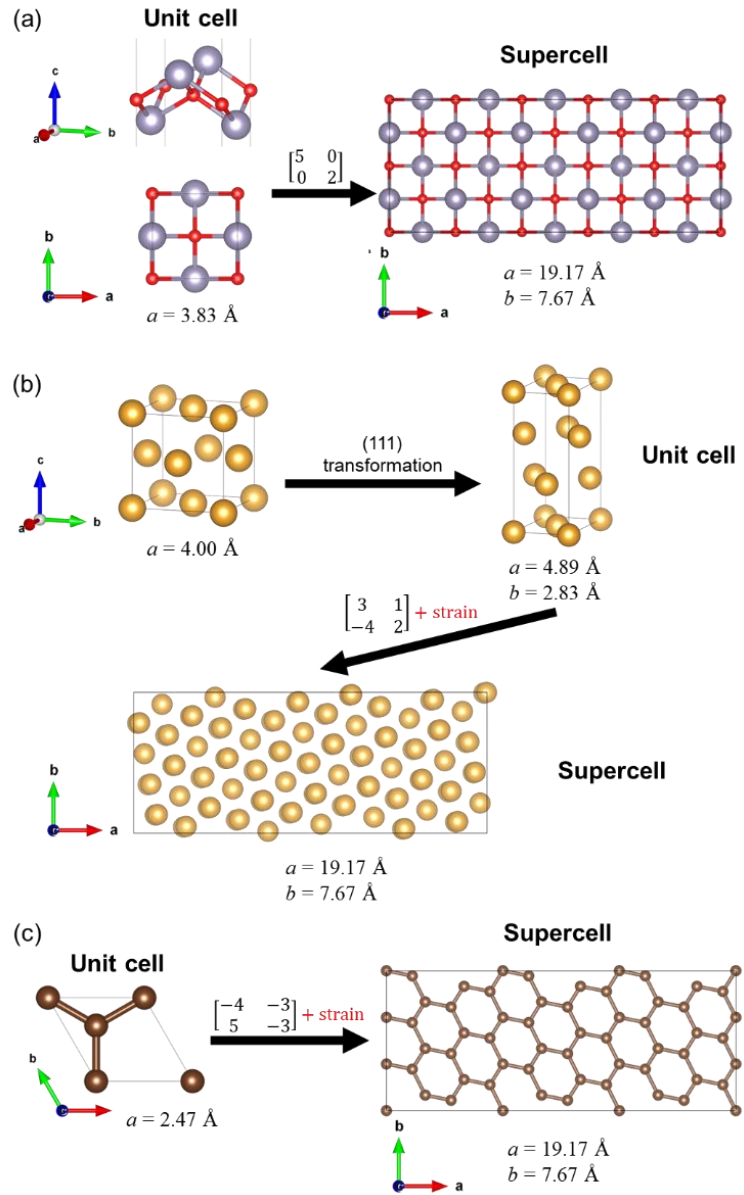

**Figure 1.** Construction of (a) mSnO, (b) (111) Au, and (c) graphene supercells from their respective unit cells with the transformation matrices labelled accordingly for lattice matching. The obtained lattice constants of gold and graphene unit cells are consistent with values in the literature [25-28].

We performed lattice matching between $\alpha$-SnO and the (111) Au unit cell with a strain limit of 5% (**Figure 1**(a) and (b)). The (111) orientation is selected as it is the most probable orientation found in experiments [15]. The gold layers are strained to match the optimised lattice constant of mSnO (3.83 Å, consistent with values in the literature [10, 29-33]) because a strain of up to a few percent does not significantly affect the electronic structure of metals at $E_f$ [34] while the electronic properties of SnO, such as its band structure, are sensitive to changes in the lattice parameters [6, 29, 35]. The solution that gives the minimum surface area was adopted to reduce the computation cost. The constructed supercells have lattice constants $a = 19.17$ Å and $b = 7.67$ Å, and each metal layer consists of 20 gold atoms.

To reduce the computation cost of the structural relaxation calculations, a step-by-step approach was adopted to construct the heterostructure supercells. An initial scanning was performed with only two metal layers by changing the interlayer distance $d$ between the relaxed gold layers and mSnO, which is taken as the difference in the average $z$ coordinates of gold and tin atoms at the interface, followed by structural relaxation *via* density functional theory (DFT) calculation. The structure with the lowest energy provides an estimate of the optimised Au–mSnO interlayer distance $d$. Gold atoms in the (111) orientation of the face-centred cubic configuration are then stacked layer by layer to this structure. The top gold layer was kept fixed in all the relaxation calculations, similar to the modelling of other metal–monolayer semiconductor structures in the literature [24, 34].

The Au–graphene–mSnO supercell is constructed as shown in **Figure 1**(c) and **Figure 2**. We first performed lattice matching between graphene and the Au–mSnO supercell by fixing the size of the latter with a strain limit of 6% for graphene. The graphene supercell with the least number of atoms is selected, with principal strains of 5.246% and −0.004%. This graphene supercell is then placed on top of fully relaxed mSnO first, followed by structural relaxation. Gold layers are then stacked on top of this structure in a similar manner with initial scanning using two layers of gold atoms followed by layer-by-layer addition. The orientation of the graphene layer is significantly altered after introducing the gold layers.

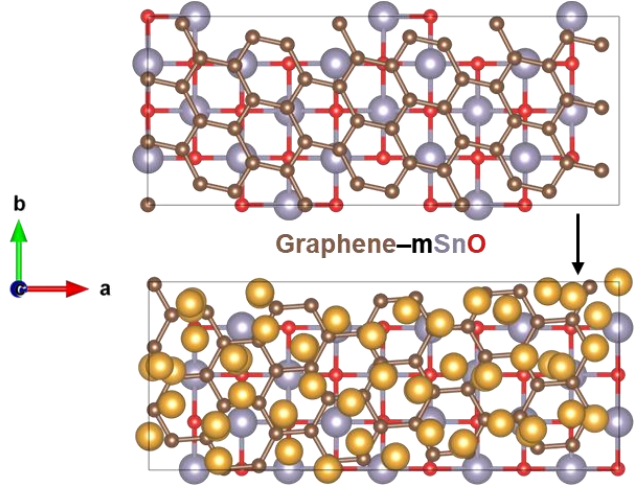

**Figure 2.** Construction of the SnO–graphene–Au supercell (top view).

**Out-of-plane compression**

To simulate out-of-plane compression, a uniaxial compressive strain is applied in the vertical direction of the heterostructures at steps of 1%, followed by structural relaxation. The compressive strain $\varepsilon$ is represented by the percentage change in the interlayer distance, and this approach has been well adopted in modelling the out-of-plane compression of heterostructures in the literature [36-38]. The corresponding stress $\sigma$ was calculated according to the Nielsen–Martin scheme [39] as

$$\sigma = \frac{\partial E}{V \partial \varepsilon}, \qquad (1)$$

where $E$ is the total strain energy, $V$ is the volume under the given compressive strain obtained as the slab area multiplied by the structure thickness.

**Density functional theory calculations**

The DFT calculations in this work were performed using the Vienna *Ab initio* Simulation Package [40] under the generalised gradient approximation (GGA) functional methods with the Perdew–Burke–Ernzerhof formulation [41]. Li *et al*. [42] showed that the DFT-D2 method of Grimme [43] can be used to obtain a relatively accurate *c*/*a* ratio of the α-SnO unit cell with respect to the experimental values at a reasonable time cost. Since the band gap of SnO has been shown to depend sensitively on its *c*/*a* ratio [35], the DFT-D2 method is selected in this work to achieve high computation accuracy.

Periodic boundary conditions are applied in the in-plane (*x* and *y*) directions. Along the normal (*z*) direction of the heterostructures, with a lattice constant of 40 Å, free boundary conditions are enforced with sufficient vacuum separation that eliminates spurious interaction between the slab structures. A kinetic energy cutoff of 500 eV is chosen for the plane wave basis set. Using the Monkhorst–Pack method, the reciprocal space is sampled with a 2 × 5 × 1 k point grid in the Brillouin zone. The energy convergence criterion is set to be $10^{-4}$ eV, and the structures are relaxed until the maximum Hellmann–Feynman force per atom is less than 0.01 eV/Å. With the relatively large supercell sizes, we adopt calculations without considering spin–orbit coupling to reduce the calculation time. Visualisations of the structures are done with the software Visualisation for Electronic and STructural Analysis (VESTA) [44].

For electronic property calculations, the energy convergence criterion is tightened to $10^{-6}$ eV. The DOS and band structures are computed for each optimised structure. The high-symmetry path X–Γ–M–R–Z–Γ for a simple tetragonal lattice is used for the band structure computation, where 20 k points are sampled in each subpath (100 in total). By comparing the energy levels of pristine mSnO and the heterostructures, the lateral SBHs can be obtained for applications in FETs.

## 3. RESULTS AND DISCUSSION

First-principles calculations are herein performed to systematically compare the tunnelling behaviour, bonding strength, charge transfer characteristics, and electronic properties of Au–mSnO and Au–graphene–mSnO structures. The results obtained from the different aspects are consistent, and they are analysed in detail to reveal the mechanism of Fermi level depinning. The changes upon out-of-plane compression of the structures are also examined.

**Tunnel barriers in the optimised heterostructures**

**Figure 3**(a) presents the optimised Au–mSnO structure, which reflects the nature of the ideal interface theoretically. Although in real situations, the contact metal consists of many layers, we restrict the calculation to only five layers of gold atoms because the interlayer distance $d$ remains relatively constant beyond a thickness of four layers of gold atoms (**Figure 3**(b)), which indicates that mSnO only interacts with the bottom metal layers. The optimised average distance $d$, which is the maximum width of the tunnel barrier between the two materials [15], for the structure with five gold layers is 2.39 Å, shorter than the sum of the atomic radii $s$ of gold and tin atoms (3.02 Å). All the tin atoms at the interface form covalent bonds with at least one gold atom as the atom-to-atom distance between the two species is shorter than $s$ with a mean value of 2.74 Å. Thus, it can be concluded that chemisorption occurs at the contact interface formed between mSnO and gold.

The insertion of graphene greatly increases the layer separation (graphene–mSnO: 3.45 Å; graphene–Au: 3.07 Å), but it is still shorter than the reported value for the Al–graphene–monolayer $MoS_2$ ($mMoS_2$) structure (~7 Å) [45]. No covalent bond is formed between graphene and either of the other two materials, and the structure becomes a van der Waals (vdW) structure, *i.e.*, physiosorption occurs instead. Note that $d$ in the optimised graphene–mSnO structure without gold is calculated to be 3.54 Å, slightly higher than that in the structure with the gold layers.

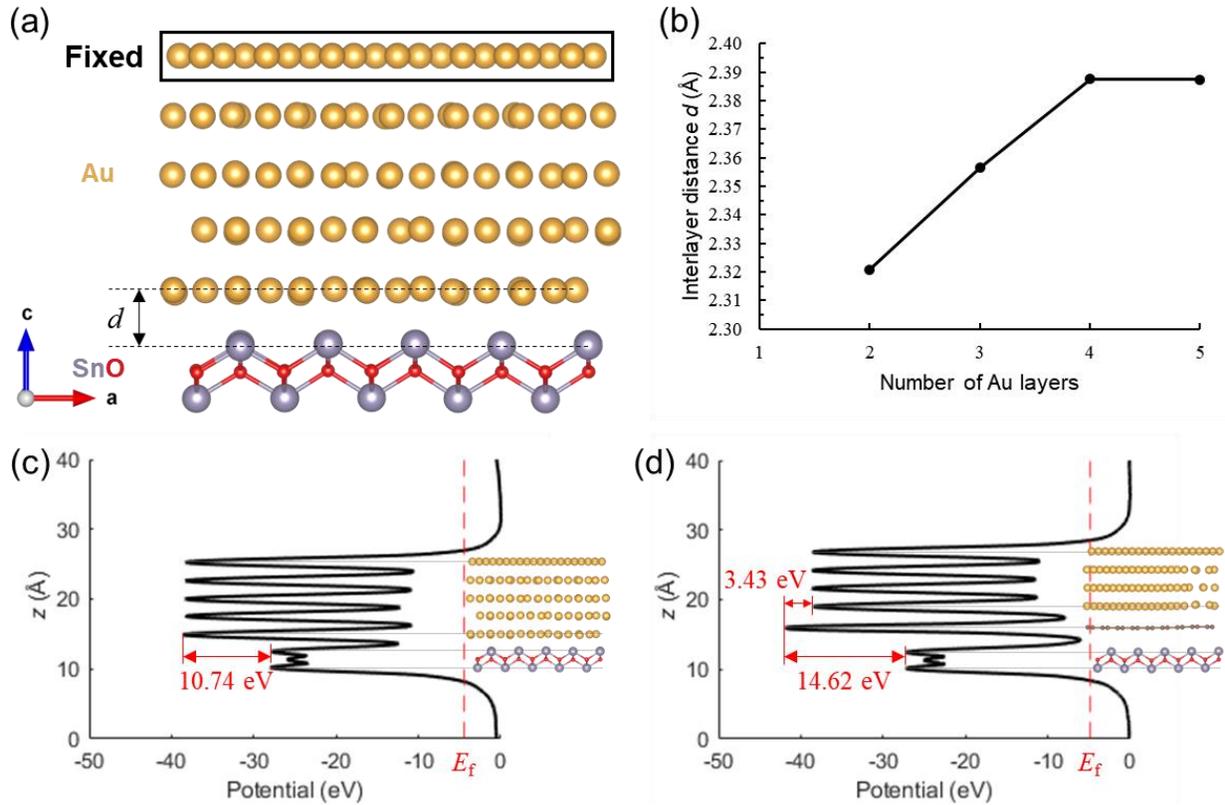

**Figure 3.** Tunnel barrier at the Au–mSnO contact with and without the graphene interlayer. (a) Optimised Au–mSnO structure with five layers of gold atoms with the interlayer distance $d$ indicating the maximum width of the tunnel barrier. (b) Changes in $d$ as the number of gold layers increases in the Au–mSnO structure. Average potential in the $xy$ plane along the $z$ direction of the (c) Au–mSnO and (d) Au–graphene–mSnO structures. The vacuum energy level is calibrated to be 0 eV, and the Fermi level $E_f$ is labelled.

**Figure 3**(c) and (d) show the tunnel barriers at the interfaces. The large potential drop of 10.74 eV from mSnO to Au indicates a strong electrical field across the interface, which may decrease the electron transfer efficiency, although the physical separation $d$ is small. After inserting graphene, this interlayer forms a much smaller tunnel barrier of 3.43 eV with gold, while the one formed with mSnO is significantly larger (14.62 eV).

**Bonding strength at the interfaces**

To evaluate the strength of the interlayer interaction forces, the interlayer binding energy $E_b$ at the Au–mSnO interface is calculated by

$$E_b = \frac{E_{Au-mSnO} - E_{Au+mSnO}}{A}, \quad (2)$$

where $E_{Au-mSnO}$ is the total energy of the hetrostructure, $E_{Au+mSnO}$ is the sum of the total energy of the isolated Au layers and mSnO fixed in the lattice of the heterostructure, and $A$ is the interface area (146.97 Å$^2$). A negative $E_b$ value indicates a stable structure, and the greater its magnitude, the stronger the binding. The calculated binding energy is −117.4 meV/Å$^2$, one order of magnitude higher than that of typical vdW heterostructures such as graphene–monolayer GeN$_3$ (mGeN$_3$) (−13.4 meV/Å$^2$; $d$ = 3.5 Å, $s$ = 1.99 Å) [46], indicating the extraordinary binding strength of the Au–mSnO heterostructure.

Gold ([Xe] 4f$^{14}$ 5d$^{10}$ 6s$^1$) has been shown to have weak adhesion with many 2D materials including graphene (presented below), phosphorene [47], bismuthene [24], and monolayer WTe$_2$ [48] as the fully filled d orbitals can not form covalent bonds. In SnO, each Sn atom ([Kr] 4d$^{10}$ 5s$^2$ 5p$^2$) shares its two 5p electrons with the neighbouring O atoms while the remaining 5s electrons are not involved in the bonding process and constitute a lone pair [30]. The high binding energy here is ascribed to the lone-pair electrons that form bonds with the gold layers. Although the bonding is strong, the structure of mSnO is shown to be almost not distorted by comparing **Figure 1**(a), **Figure 2**, and **Figure 3**(a).

To determine the suitable order of experimentally fabricating the layers in the Au–graphene–mSnO structure, the binding energy at the Au–graphene and graphene–mSnO interfaces are similarly calculated (−62.3 and −14.1 meV/Å$^2$, respectively). The negative values confirm the stability of the structures. The smaller magnitude than that of the Au–mSnO structure is attributed to the weaker vdW bonding than covalent bonding. The Au–graphene substructure has much bigger magnitude than both the graphene–mSnO substructure here and graphene–mGeN$_3$ structure quoted above, indicating that it is a rather stable vdW structure and hence should be fabricated first. The binding energy of the interface formed between the fabricated Au–graphene substructure and mSnO is computed to be −21.7 meV/Å$^2$. Its negative value and relatively big magnitude confirm the feasibility of this fabrication route.

**Charge transfer across the interfaces**

**Table 1** displays the computed work functions, and those of the pristine materials are consistent with values in the literature [49-51]. The pristine Au (111) surface exhibits a slightly lower work function than mSnO. Since mSnO is a p-type semiconductor, Schottky contact is expected at the Au–mSnO interface by the classical Schottky–Mott rule [52, 53]. Electrons should be transferred from gold to mSnO to equilibrate the Fermi levels.

**Table 1.** Work functions of the fully relaxed materials systems in this work.

| Materials system | Work function (eV) |
|---|---|
| Pristine mSnO | 5.16 |
| Pristine Au (111) surface | 5.02 |
| Au–mSnO | 4.52 |
| Au–graphene–mSnO | 4.83 |

The Fermi level shift $\Delta E_f$ can be defined as [47]

$$\Delta E_f = W - W_p, \qquad (3)$$

where $W$ and $W_p$ are the work functions of the heterostructure and pristine mSnO, respectively. It has a value of −0.64 eV for the Au–mSnO structure, and the insertion of graphene decreases its magnitude ($\Delta E_\text{f} = -0.33$ eV). This decrease is due to the weaker interaction of mSnO with graphene than with Au, as indicated by the smaller binding energy magnitude at the former interface as presented above. The negative $\Delta E_\text{f}$ values of both structures indicate that electrons are transferred from gold to mSnO, consistent with the prediction by the Schottky–Mott rule.

To understand the charge transfer at the Au–mSnO interface, the charge density difference $\Delta\rho$ is calculated as

$$\Delta\rho = \rho_\text{Au-mSnO} - \rho_\text{Au} - \rho_\text{mSnO}, \quad (4)$$

where $\rho_\text{Au-mSnO}$, $\rho_\text{Au}$, and $\rho_\text{mSnO}$ represent the charge density of the Au–mSnO structure, Au layers, and mSnO. The charge transfer at the Au–graphene–mSnO contact is also similarly calculated.

As shown in **Figure 4**(a), in the Au–mSnO structure, charge redistribution mainly occurs at the interface between mSnO and the bottom two layers of gold atoms, similar to the case of metal–phosphorene structures [47]. The electron accumulation between the gold and tin atoms highlights that the nature of the bonding between them is covalent. Such electron accumulation has also been observed in a Mg–mMoS$_2$ structure at its optimised interlayer distance $d = 2.2$ Å, which disappears at $d = 5$ Å where the covalent bonds between the magnesium and sulphur atoms are expected to break with $s = 2.64$ Å [19].

The s orbital of Sn becomes depleted in the Au–mSnO structure as indicated by the cyan colour, and it hybridises with the orbitals of the gold atoms to form covalent bonds. The strong orbital overlap may be attributed to the small lattice mismatch between the two materials [15]. The accumulation of electrons from occupied gap states and bulk gold at the surface of gold results in a decrease in its work function, *i.e.*, $E_\text{f}$ is increased and pinned at the upper region of the mSnO band gap, leading to an n-type character at the Au–mSnO interface. Being electronegative, the oxygen atoms in mSnO pull electrons towards themselves.

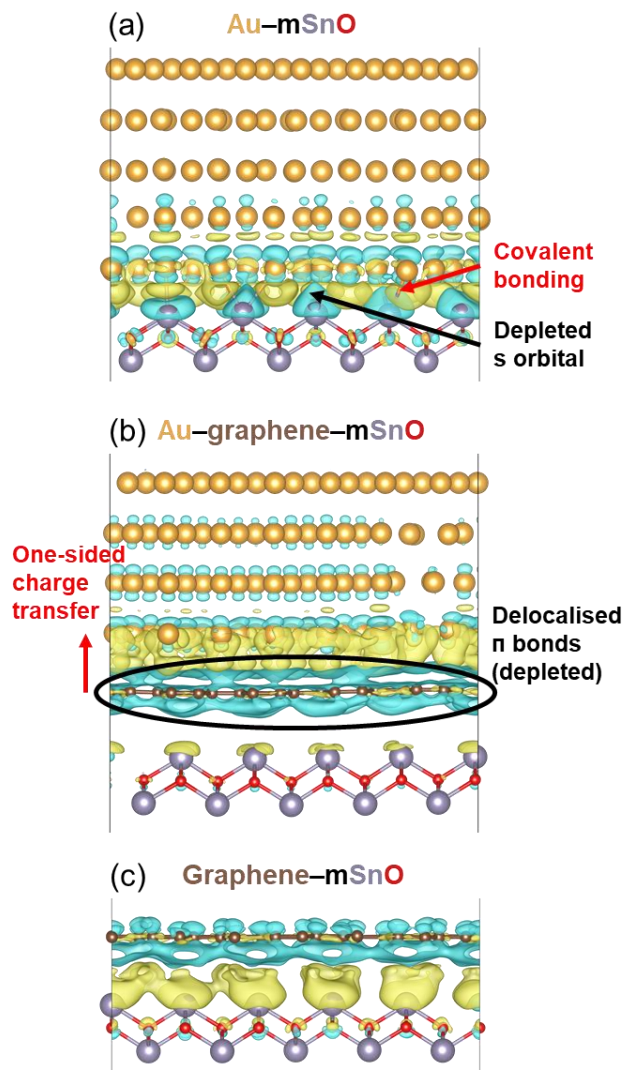

**Figure 4.** Three-dimensional isosurface of the charge density difference. (a) Au–mSnO (isosurface: 0.0015 e/Bohr$^3$). (b) Au–graphene–mSnO (isosurface: 0.0003 e/Bohr$^3$). (c) Graphene–mSnO (isosurface: 0.0001 e/Bohr$^3$). Cyan: electron depletion; yellow: electron accumulation.

**Figure 4**(b) shows that after inserting the graphene layer, the mSnO bands are much less perturbed by the interfacial interaction, especially considering the isosurface level is five times lower, indicating the weak bonding between mSnO and graphene. Instead of electron sharing in the form of covalent bonds, electrons are transferred from the delocalised π bonds in the graphene layer to mainly the bottom gold layer and much less so to mSnO, thereby creating an interface

dipole. This one-sided charge transfer could be attributed to the significantly lower tunnel barrier between graphene and gold than the one between graphene and mSnO as shown above. **Figure 4**(c) demonstrates that when there are no gold layers, electrons will be transferred from graphene to mSnO only, without forming covalent bonds because of the large interlayer distance. Note that the isosurface level here is even smaller, indicating the much weaker interaction.

**Electronic properties**

This section compares the different electronic properties, including the band structure (**Figure 5**), density of states (**Figure 6**), and Schottky barriers (**Figure 7**), of the structures with and without the graphene layer.

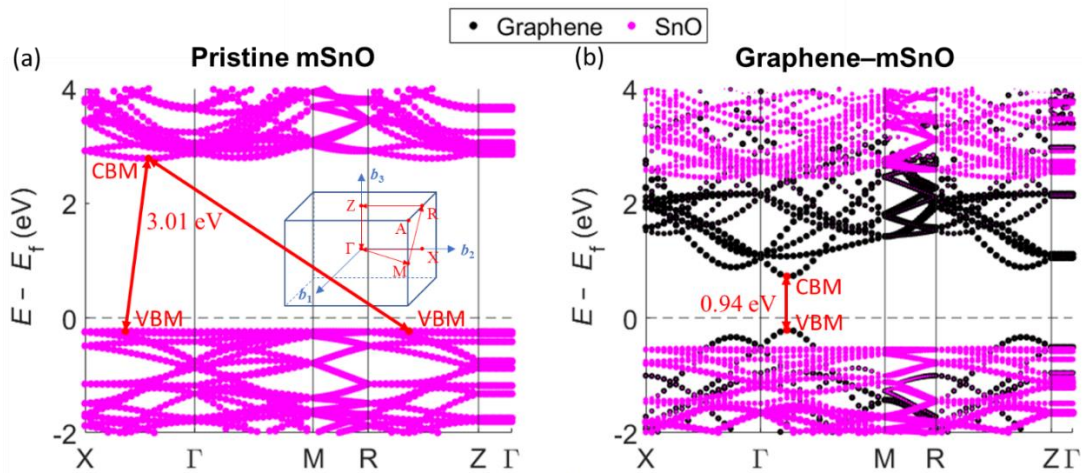

**Figure 5.** Band structures obtained using the GGA method with a high-symmetry path X–Γ–M–R–Z–Γ. (a) Pristine mSnO supercell with an indirect band gap of 3.01 eV. The inset shows the simple tetragonal lattice containing the path. (b) Graphene–mSnO with a direct band gap of 0.94 eV. The dot size is proportional to the weight of the states' projection. Band gap = conduction band minimum (CBM) – valence band maximum (VBM).

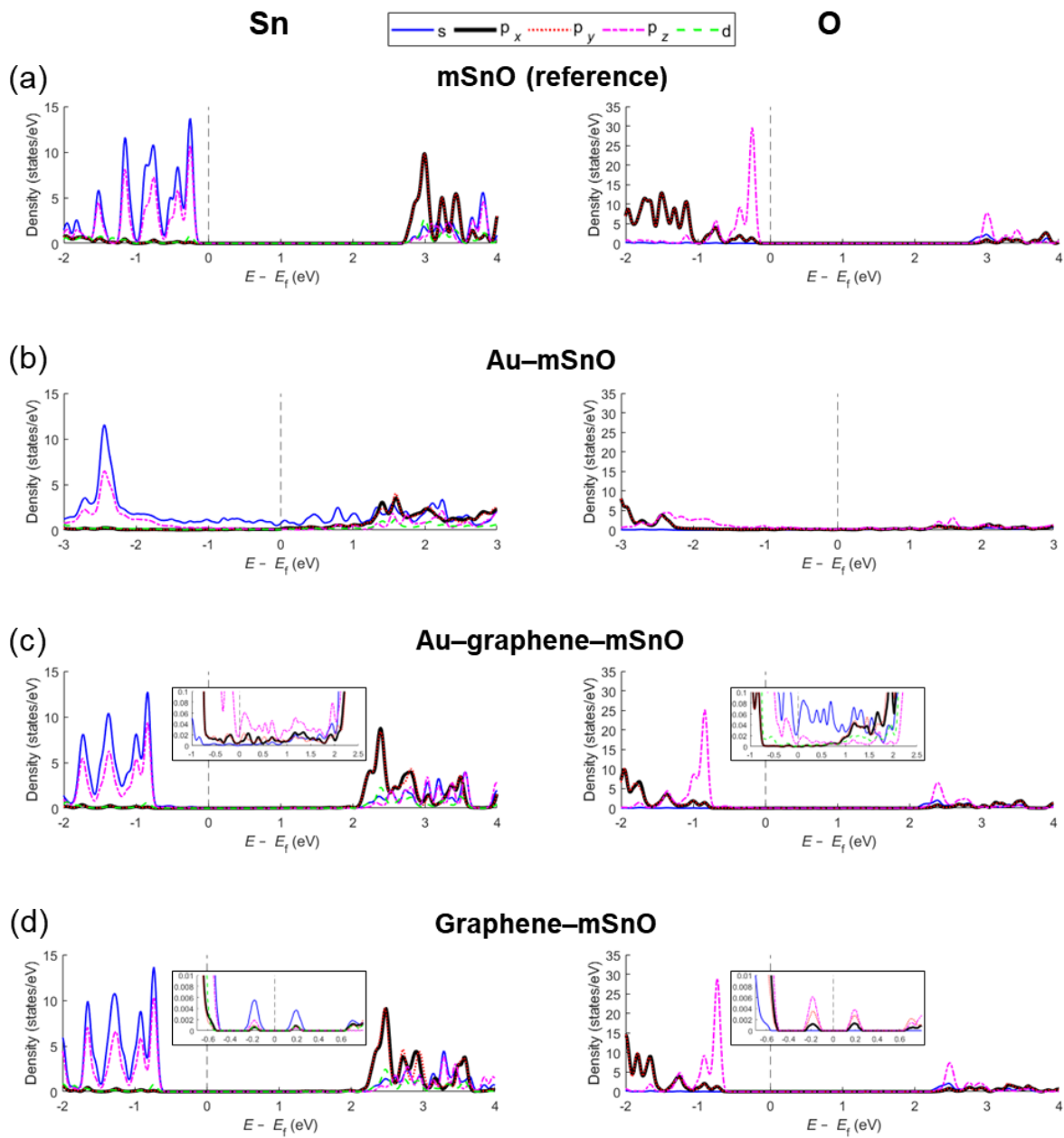

**Figure 6.** DOS projected on Sn and O orbitals in different structures. (a) mSnO supercell. (b) Au–mSnO. (c) Au–graphene–mSnO. (d) Graphene–mSnO. The insets are zoomed-in plots for the energy range of −1–2.5 eV in (c) and −0.7–0.8 eV in (d), showing the non-zero states near the Fermi level $E_f$.

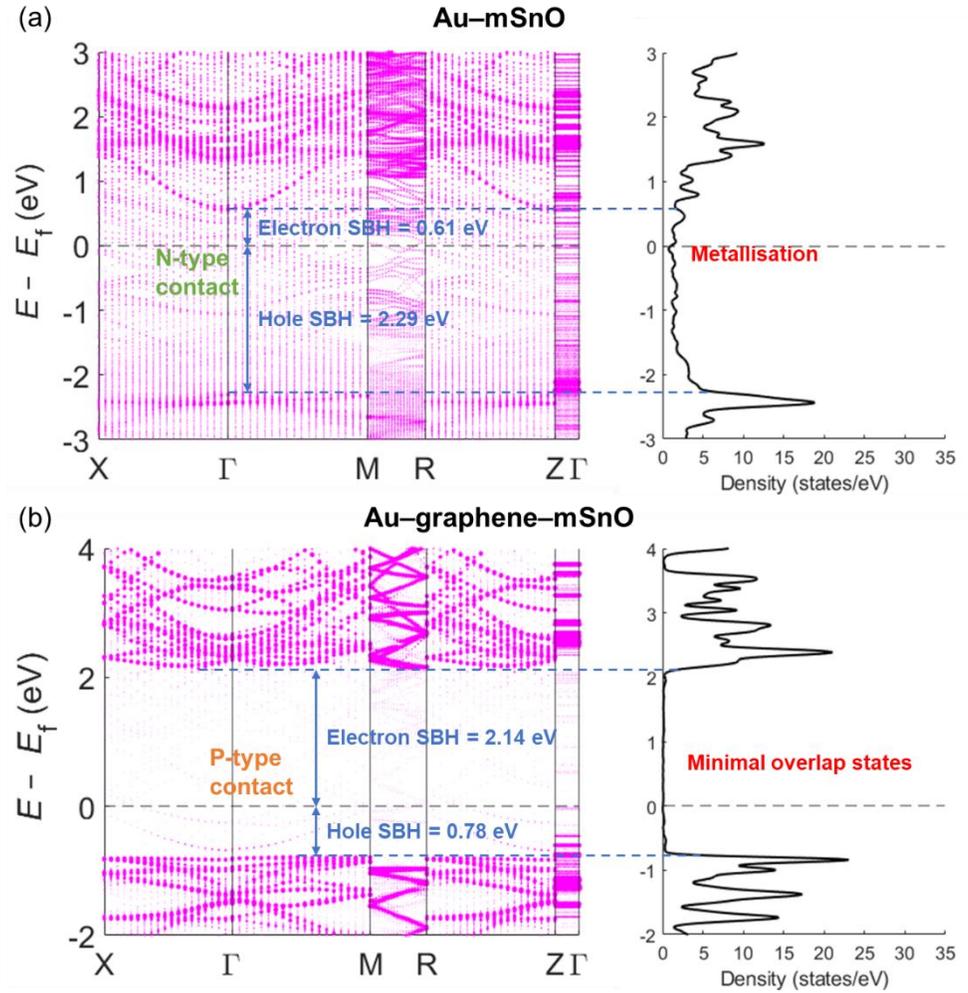

**Figure 7.** Projected band structure (magenta dots of sizes proportional to the weight of the states' projection) and DOS on SnO orbitals for (a) Au–mSnO and (b) Au–graphene–mSnO. The Fermi level $E_f$ is aligned to determine the positions of CBM and VBM.

Pristine mSnO is a semiconductor. The pristine monolayer used in our heterostructures has an indirect band gap of 3.01 eV (**Figure 5**(a) and **Figure 6**(a)), similar to the literature values obtained at the GGA level [30]. In the Au–mSnO structure, the strong band hybridisation due to the covalent bonds between the two materials significantly distorts the properties of mSnO as a semiconductor, as shown by the MIGS filled in the original band gap of mSnO (**Figure 6**(b) and **Figure 7**(a)). As the monolayer no longer exhibits a band gap, it undergoes metallisation. Vanishing of the band gap of a 2D semiconductor in top contact with a metal *via* covalent bonds has also been reported in the literature [15, 24].

The conduction and valence bands are not easy to identify because of the strong hybridisation. The positions of the CBM and VBM can be estimated by locating the corresponding curvature with significant weight of the states' projection [18, 54], and lines can be drawn from these positions extending to the projected DOS (PDOS) plot placed beside with $E_f$ aligned. Within the interval confined by these two lines, the MIGS contribution in the DOS is minimal, which confirms the estimated positions. The electron and hole SBHs are estimated to be 0.61 and 2.29 eV, respectively. Because of the significant increase in $E_f$, the electron SBH is much lower than the hole SBH, resulting in n-type contact, which is consistent with the analysis above on charge transfer. As shown in **Figure 6**(b), the MIGS in the Au–mSnO structure are mainly s states of tin as the lone-pair s electrons are pulled to form the covalent bonds, and they are the main driving force that strongly pins $E_f$ in the band gap of mSnO.

Since the monolayer is metallised, it may form a lateral SB with the mSnO channel when it is used in FETs (**Figure 8**). We estimate the SBH by the work function approximation approach, where the coupling between regions B and C is ignored [24, 47, 48]. The lateral hole (electron) SBH, which is calculated as the difference between the Fermi level of the metallised monolayer (B) and the VBM (CBM) of the free-standing channel material (C, pristine mSnO), is 0.87 (2.14) eV. A lower lateral hole SBH is obtained, indicating that p-type Schottky contact is formed at interface II.

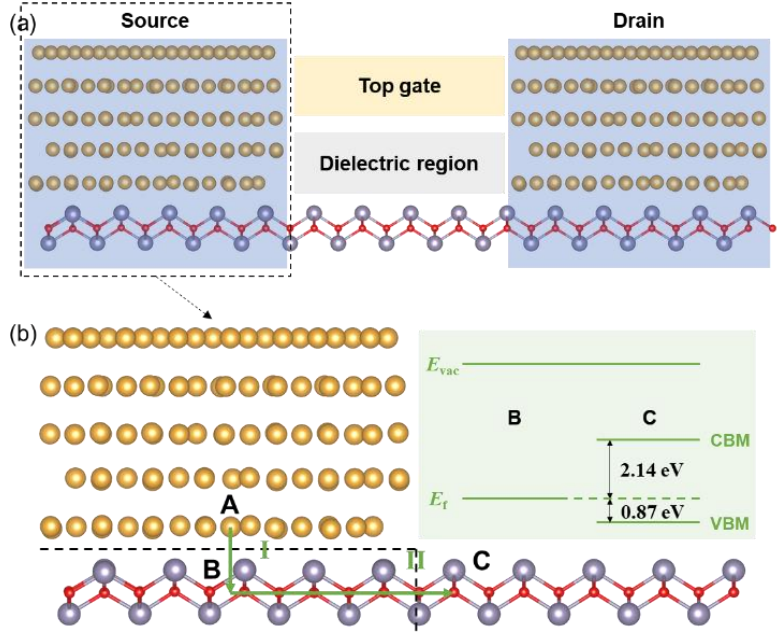

**Figure 8.** (a) Cross-sectional view of an mSnO FET. (b) Metal contact to the mSnO channel. A, B, and C present three regions, while I and II denote the interfaces between these regions. The green arrows show the pathway (A → B → C) of charges from the contact metal (A) to the channel. The energy levels of regions B and C are shown on the right with $E_{\text{vac}}$ denoting the vacuum energy level. The electron and hole SBs are labelled as the energy differences between the Fermi level $E_f$ of region B, and the CBM and VBM, respectively, of region C.

As indicated by the faint band states around $E_f$ in **Figure 7**(b), the MIGS are almost destroyed, owing to the inserted graphene layer breaking the interfacial interaction. The much fewer MIGS are considered overlap states [13, 15, 48, 55]. Neglecting their minimal contribution in the DOS, we determine the band gap to be 2.92 eV, lower than the one of pristine mSnO. Since the hole SBH (0.78 eV) is smaller than the electron SBH (2.14 eV), the contact is p-type, which is surprising given that gold has a rather high work function. Compared to a similar case of a graphene interlayer at the Ag−mMoS$_2$ contact that decreases the SBH from 0.3 to 0.19 eV [56], the changes here due to graphene insertion is much more significant. The lateral hole and electron SBHs at interface II are computed to be 0.56 and 2.45 eV, respectively. The lateral hole SBH slightly decreases after the insertion, indicating a smaller contact resistance.

**Figure 6**(a) and (c) demonstrate that the PDOS curves of mSnO in the structure with the graphene interlayer are almost the same as those of the pristine monolayer, which could be attributed to the one-direction charge transfer from graphene to gold instead of mSnO that prevents significant perturbance to the energy of mSnO orbitals. As shown by the insets in **Figure 6**(c), the PDOS around $E_f$ is non-zero but close to zero, indicating that the monolayer experiences much weaker metallisation than that without the interlayer.

**Figure 5**(b) shows that without the gold layers, the graphene–mSnO structure exhibits a direct band gap of 0.94 eV. The much bigger black dots near the CBM and VBM indicate that the smaller band gap than the one of the pristine mSnO is mainly attributed to the graphene states. The insets in **Figure 6**(d) display small humps within the band gap of the original monolayer, which belong to the graphene-induced gap states. As the PDOS plots capture states that are closer to $E_f$ than the band structure plot, the band gap is recalculated to be only 0.05 eV.

Although graphene can be considered a metal with no band gap, when it is in contact with mSnO alone, the semiconducting properties of the latter can be preserved. It can then be concluded that the MIGS in the Au–graphene–mSnO structure are mainly due to gold, and graphene acts as a successful buffer layer that shields and preserves the semiconductor nature of mSnO to a large extent. The SnO monolayer is much less metallised than the case without graphene. The much fewer states near $E_f$ may not accommodate many electrons.

Note that all the results obtained in this work are based on the assumption of perfect interfaces, which requires the removal or prevention of surface impurities and annealing [57]. Contamination of the semiconductor surface may lead to surface reactions and cause even stronger FLP.

**Out-of-plane compression**

During the operation, test, and transport of electronic devices, they are prone to mechanical deformation, and strain has been shown to significantly affect the electronic properties of semiconductors [29, 58-60]. For example, the touchdown of pogo pins exerts out-of-plane compression to the structure. For layered materials with interlaminar forces, the interlayer distance is decreased under out-of-plane compression. To verify the performance of the investigated heterostructures under such deformation, its effects on the electronic behaviours are studied here by decreasing the interlayer distance.

**Figure 9**(a) shows the obtained stress–strain plot for the Au–mSnO structure. The increase in stress becomes more significant as the structure is compressed further. The typical out-of-plane compressive stress during touchdown of pogo pins ranges from 247 to 390 MPa, which is within the stress range investigated in this study, corresponding to up to 3% strain. **Figure 9**(b) shows that for the Au–graphene–mSnO structure, the changes in the individual interlayer distances are not the same. The greater changes in the graphene–mSnO layer separation could be attributed to the weaker interaction between these two materials (**Figure 4**(b)) that provides lower resistance to compression. This finding is also corroborated by the fact that the magnitude of the binding energy is smaller at the graphene–mSnO interface than at the Au–graphene interface as presented above.

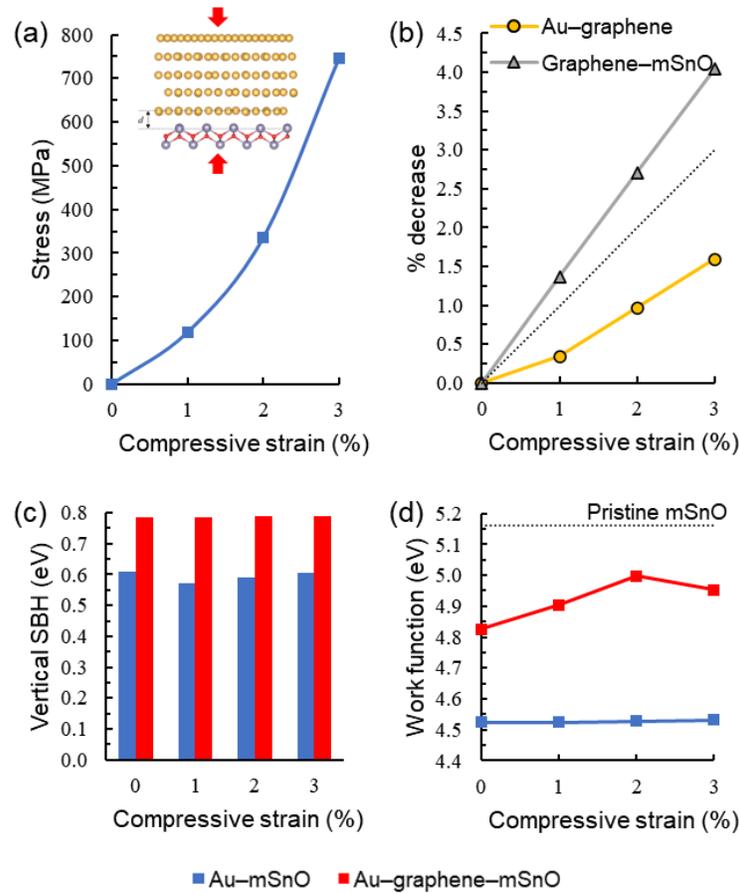

**Figure 9.** Effects of out-of-plane compression on the Au–mSnO structure with and without the graphene interlayer. (a) Stress–strain curve of the Au–mSnO structure. (b) Percentage decrease in the Au–graphene and graphene–mSnO layer separation of the Au–graphene–mSnO structure with the 45° line as a reference. (c) Vertical SBH and (d) work function of Au–mSnO (blue) and Au–graphene–mSnO (red). The work function of pristine mSnO is indicated by the black dotted line for reference.

The vertical SBHs of both structures remain stable under the out-of-plane compression, especially for the structure with the graphene interlayer (**Figure 9**(c), **Figure 10**, and **Figure 11**), indicating the consistent charge transfer performance in the vertical direction during such deformation. The work function is relatively constant for the Au–mSnO structure under the out-of-plane compression (**Figure 9**(d)) as small compression does not significantly change the amount of charge transferred across the interface bonded by covalent forces. In contrast, the work function becomes rather sensitive to such small compression when graphene is inserted as the efficiency of charge transfer is much improved with a smaller vdW gap. The significant changes in the work function indicate successful Fermi level depinning. A higher work function lowers the lateral hole SBH, and insertion of graphene would hence allow for tunability of the charge transfer into the channel material *via* out-of-plane compression.

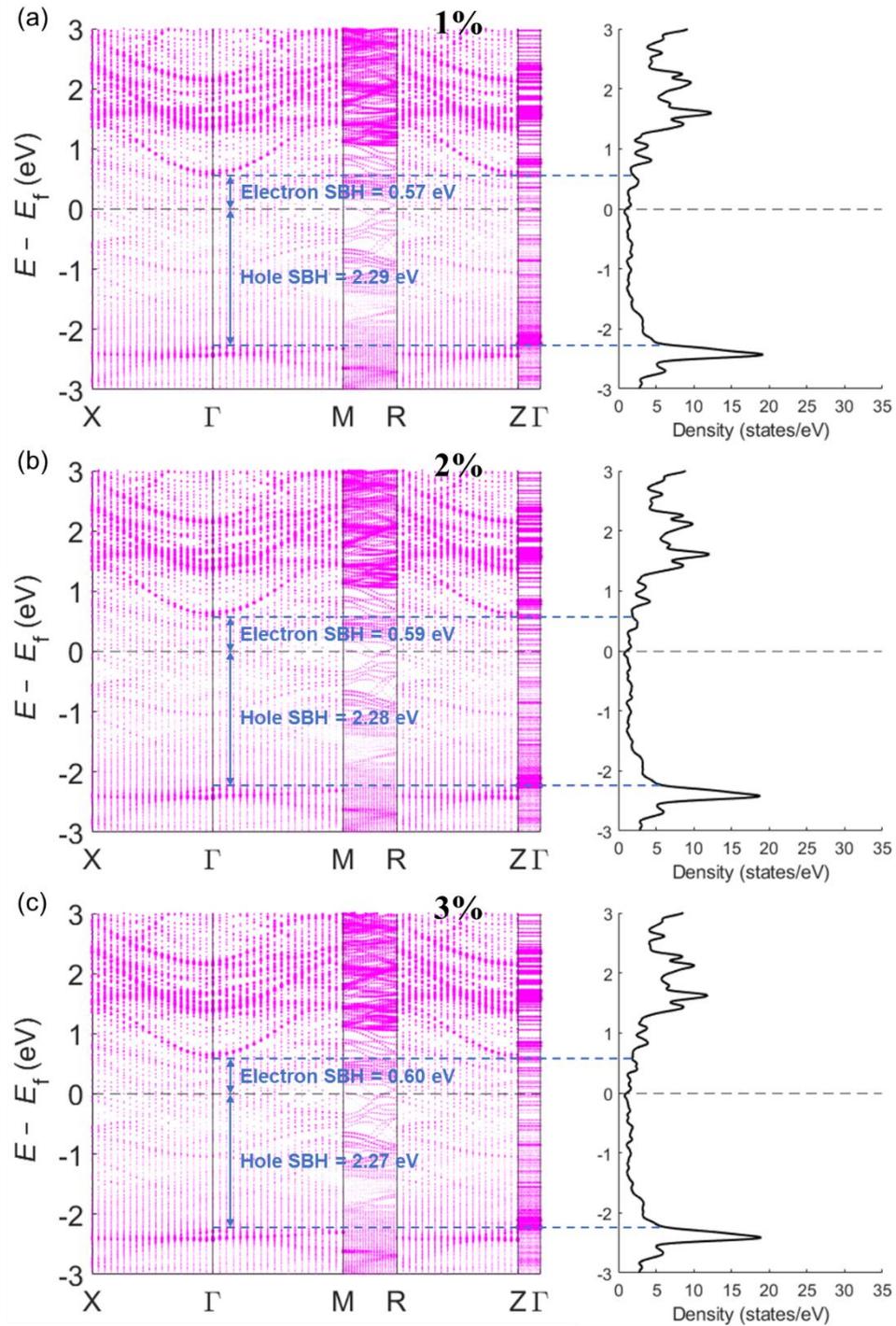

**Figure 10.** Projected band structure (magenta dots of sizes proportional to the weight of the states' projection) and DOS on SnO orbitals for Au–mSnO under out-of-plane compression of (a) 1%, (b) 2%, and (c) 3%. The Fermi level $E_f$ is aligned to determine the positions of CBM and VBM.

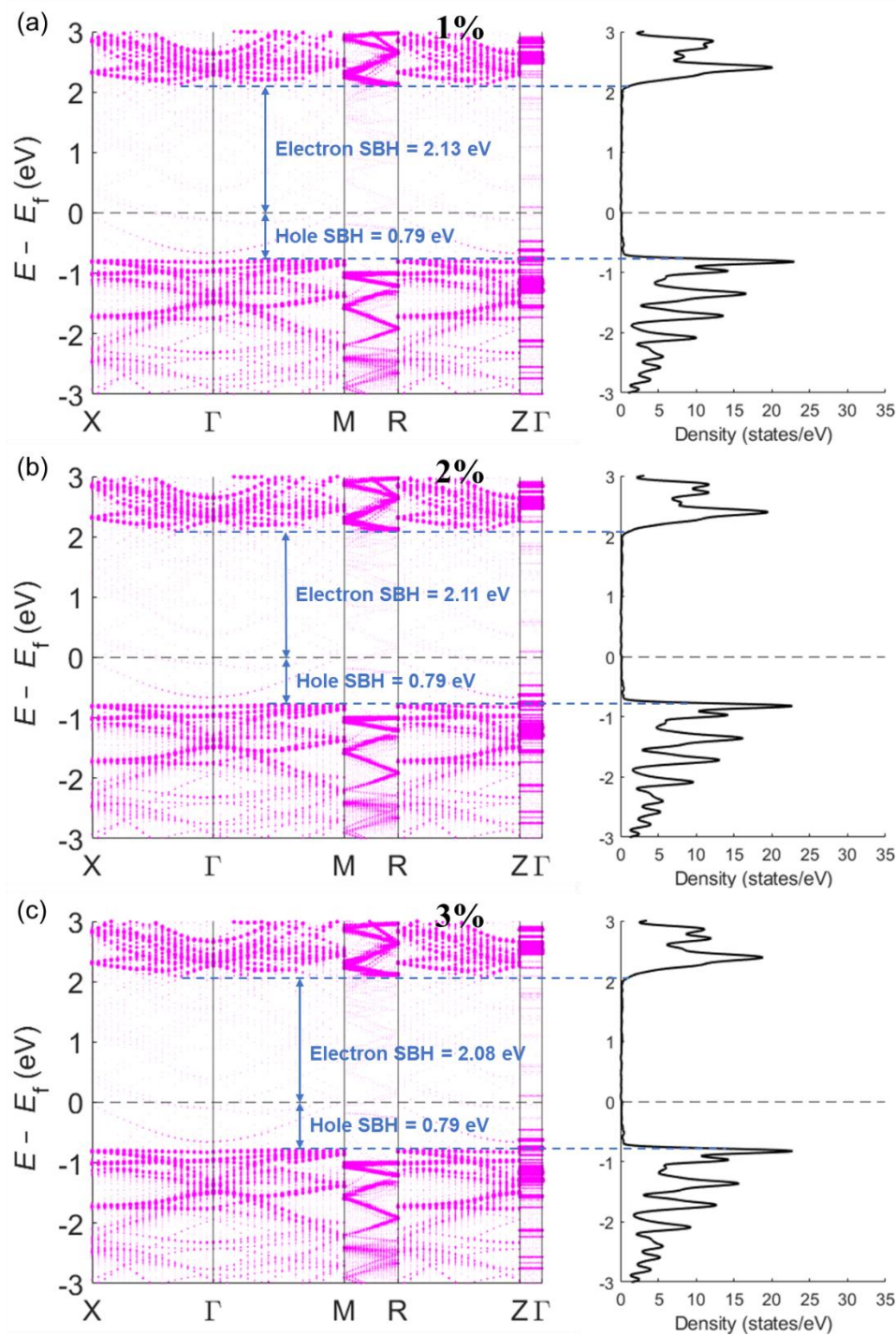

**Figure 11.** Projected band structure (magenta dots of sizes proportional to the weight of the states' projection) and DOS on SnO orbitals for Au–graphene–mSnO under out-of-plane compression of (a) 1%, (b) 2%, and (c) 3%. The Fermi level $E_f$ is aligned to determine the positions of CBM and VBM.

# 4. CONCLUSION

We performed systematic analysis on the electronic contact between gold and mSnO and its changes when graphene is inserted between them. The close examination of the geometry, binding energy, and charge transfer demonstrates that strong covalent interaction is formed between gold and mSnO although gold has fully occupied d orbitals. Such strong bonding is attributed to the lone-pair electrons of tin atoms in SnO, and it has been demonstrated to contribute to the FLP effect as indicated by the MIGS filled in the original band gap of mSnO. The insertion of graphene helps to depin $E_f$ by forming weak vdW interactions instead of covalent bonds. Charges are transferred from graphene to mainly gold but not mSnO because of the significantly lower tunnel barrier between the former two materials. This one-sided charge transfer causes less perturbance to the electronic structure of mSnO, hence preserving its semiconducting properties to a large extent.

To facilitate the application of mSnO in monolayer FETs, we computed and compared the vertical and lateral SBHs at the Au–mSnO contact with and without the graphene interlayer. Because of the FLP effect, n-type vertical contact is formed between gold and mSnO, while the graphene insertion surprisingly leads to p-type vertical contact, given that gold already has a high work function. P-type lateral contact is formed regardless of the presence of the interlayer, but with graphene, the hole SBH is lower, indicating a smaller contact resistance. We also investigated the effects of out-of-plane compression (up to 3%) to simulate touchdown of pogo pins. As such small compression may not significantly change the amount of charge transferred across covalently bonded interfaces but reduce the vdW gap, the work function of Au–graphene–mSnO structure is affected more significantly.

This work scrutinises the possibilities of employing a promising 2D semiconductor, mSnO in monolayer FETs by systematically examining its contact with a common electrode material, gold and demonstrates successful Fermi level depinning by graphene insertion. It also guides the future exploration of 2D materials for various electronic applications and selection of buffer layers to mitigate the effects of MIGS.

# ACKNOWLEDGMENTS

This study was supported by the Economic Development Board - Singapore and Infineon Technologies Asia Pacific Pte. Ltd. through the Industrial Postgraduate Programme with Nanyang Technological University (NTU). The computational calculations for this study were partially conducted using the resources of the National Supercomputing Centre, Singapore and High-Performance Computing Centre, NTU.